\documentclass[%
 reprint,
superscriptaddress,
nofootinbib,
nobibnotes,
 amsmath,amssymb,
 aps,
prb,
longbibliography
]{revtex4-2}

\usepackage{graphicx}% Include figure files
\usepackage{dcolumn}% Align table columns on decimal point
\usepackage{bm}% bold math
\usepackage[colorlinks=true,citecolor=blue,linkcolor=blue, urlcolor=cyan]{hyperref} 
\usepackage{empheq}
\usepackage{xcolor}
\usepackage[normalem]{ulem}
\usepackage[english]{babel}

\usepackage[T2A]{fontenc}
\usepackage[utf8]{inputenc}

\usepackage{orcidlink}

\begin{document}

\preprint{-}

\title{Bistability of electron temperature in atomically thin semiconductors in the presence of exciton photogeneration}

\author{A. M. Shentsev\orcidlink{0009-0003-7426-3479}}
\affiliation{Moscow Institute of Physics and Technology, 141700 Dolgoprudny, Russia}
\affiliation{L. D. Landau Institute for Theoretical Physics, 142432 Chernogolovka, Russia}

\date{\today}

\begin{abstract} 
We study the dynamic equilibrium between trions and excitons in monolayers of transition metal dichalcogenides in the presence of resident charge carriers and continuous photogeneration of excitons. We show that heating of the system via Drude absorption of low-frequency radiation leads to bistability of the steady-state equilibrium. The first is a low-temperature state, in which almost all resident charge carriers are bound into trions. The second state occurs at high temperatures, where most trions are dissociated; in this regime, the heating is more efficient due to the higher Drude conductivity of unbound charge carriers compared to trions. Switching between these two states occurs on a timescale of tens to hundreds of picoseconds and is accompanied by a jump in various observables such as temperature, current, and the intensity of exciton or trion luminescence.
\end{abstract}

\maketitle

\section{\label{sec:intro}Introduction}
Transition metal dichalcogenide monolayers (TMDC MLs) are direct-bandgap semiconductors~\cite{mak2010atomically,splendiani2010emerging} whose optical properties are largely determined by various multiparticle Coulomb complexes~\cite{Durnev:2018, ivchenko2005optical, Semina_2022} with high binding energies. These include neutral excitons (X) and charged excitons (trions): specifically, three-particle complexes in which an exciton is bound to either an electron from the conduction band (the negative trion, $X^-$) or a hole from the valence band (the positive trion, $X^+$). The exciton binding energies in these materials are on the order of hundreds of meV, while the trion binding energies are approximately 20–30 meV, i.e., the are about an order of magnitude smaller~\cite{courtade2017charged, Mak:2013lh}. Thus, due to their high stability, it is possible to study these complexes over a wide temperature range~\cite{xiao2017excitons,chernikov2014exciton, ross2013electrical,massignan2025polarons,davila2024temperature,zhang2015experimental}.

The binding energy of trions in two-dimensional (2D) TMDCs falls within the terahertz spectral range—a frequency region that is presently the focus of intense research~\cite{ganichev2006intense,Lampin2020,helm2023elbe,Lu2024,leinss2008terahertz}. Recent experimental works~\cite{venanzi2024ultrafast, cuccu2025terahertz} have demonstrated the conversion of trions into excitons, as well as the possibility of controlling the ratio of trion to exciton concentration by applying picosecond terahertz pulses. Alternatively, such conversion can occur through heating of the electron gas~\cite{venanzi2021terahertz, Budkin:2011aa}, which leads to trion dissociation via collisions with hot electrons whose energies are on the order of the trion binding energy~\cite{shentsev2025terahertz}. This collision-induced trion-to-exciton conversion, also termed impact dissociation, is more efficient at low-frequency radiation due to greater heating of the sample. The reverse process—the relaxation of excitons into trions—is most likely mediated by the emission of an optical phonon into the lattice~\cite{ayari2020phonon}. Notably, in TMDC MLs, the energy of optical phonons can be very close to the trion binding energy~\cite{PhysRevB.90.045422}; therefore, this relaxation rate is only weakly dependent on temperature~\cite{ayari2020phonon}, %in contrast to the exponentially activated process of impact trion dissociation.
in contrast to the impact dissociation of trions, which increases exponentially with temperature. 

In this paper, we investigate nonlinear effects~\cite{ashkinadze1987self,Ashkinadze:1988aa,bonilla2024nonlinear,mourzidis2025exciton,braunstein1962nonlinear,genco2022optical,wieczorek2009bifurcations} arising from the conversion of excitons to trions in TMDC MLs under continuous exciton photogeneration. Namely, the phenomenon of bistability~\cite{blanchard2006differential}, which occurs in a wide range of systems~\cite{gibbs1979optical, loth2012bistability,bory2015unipolar,semina2022valley,l4tq-bxq9, ostrovsky2006proximity, valagiannopoulos2022multistability,PhysRevLett.106.105503}, from classical oscillators to metasurfaces~\cite{diallo2026metamaterials}. We show that the exponential increase in trion dissociation rate with temperature leads to a stepwise change in the ratio of the density of unbound resident charge carriers to the trion density. Combined with the fact that trions (being heavier than free carriers) exhibit lower conductivity and shorter energy relaxation times~\cite{PhysRevB.105.075311}, this effect gives rise to bistability in the system’s steady state when it is heated by an external low-frequency field. The first state is a low-temperature one, in which most of the resident charge carriers are bound to excitons, forming trions. The second is a high-temperature state, in which most trions are dissociated, leading to more efficient heating. Switching between these two states occurs on a timescale of 10–100 ps and is accompanied by abrupt changes in various system observables—such as temperature, electric current, fraction of the absorbed-transmitted external electromagnetic field and the intensity of exciton or trion luminescence—forming a hysteresis loop as the heating power is varied. 

The article is organized as follows. In Sec.~\ref{sec:eTDyn}, we present the dynamic equations for the components of the electron gas and determine the dependence of their equilibrium densities on temperature. The differences between the absorption and cooling rate of trions and resident charge carriers are considered in Sec.~\ref{sec:HC}. In Sec.~\ref{sec:Bif}, we demonstrate the emergence of bistability in the steady state and the formation of a hysteresis loop for the system's observables. Here we also consider the influence of system parameters on the bifurcation region. 
The case of low resident charge carrier densities, as well as the process of temporal relaxation during switching between regimes,
%The process of temporal relaxation during switching between regimes, as well as the case of low resident charge carrier densities, 
is presented in Sec.~\ref{sec:Lowdenc} and \ref{sec:Switch}, respectively. The main results and conclusions are presented in Sec.~\ref{sec:Concl}.
\newpage
\section{\label{sec:eTDyn}Dynamic equations of the system}

We consider $n$-doped TMDC MLs, with an resident electron density of $n_c = 10^{11}...10^{12}$ cm$^{-2}$ in the conduction band. While the type of resident charge carrier is not crucial (and the case of $p$-doping with holes in the valence band is analogous to our consideration) we focus here on the $n$-doping scenario. Following the photogeneration of an exciton density  $n_X$, their energy relaxation by optical phonons leads to the formation of $X^-$ trions, whose density we denote
as $n_T$. At the considered resident charge carrier densities, the Fermi energy is much smaller than the trion binding energy, ensuring that the trions remain stable. 

\begin{figure*}[t]
    \centering    \includegraphics[width=\textwidth]{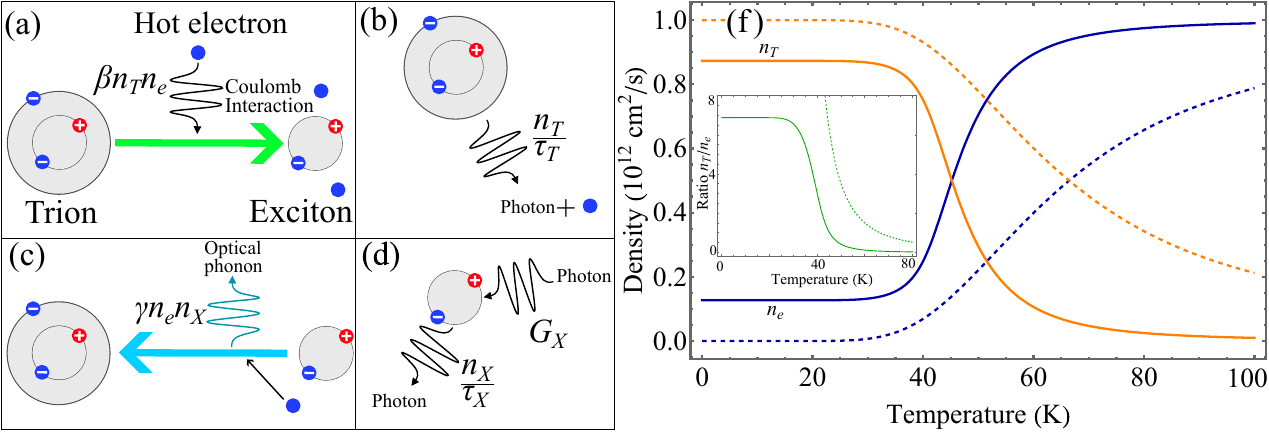}
    \caption{Diagram of the processes described by Eq.~\eqref{syst}: (a) Impact trion dissociation with the rate $\beta n_T n_e$ caused by electrons with an energy on the order of $E_T$; (b) Radiative recombination of an exciton within a trion ($n_T/\tau_T$), during which an optical photon is emitted and the outor electron remains in the conduction band; (c) Exciton energy relaxation via electron capture from the conduction band, accompanied by optical phonon emission ($\gamma n_e n_X$); (d) Exciton photogeneration $G_X$ and radiative recombination ($n_X/\tau_X$), see text for details. (f) The dependence of the density of unbound electrons $n_e$ and trion density $n_T$ on the temperature of the electron gas. The solid lines correspond to the solution of Eq.~\eqref{qudeq}, while the dashed lines correspond to the Saha equation Eq.~\eqref{eq:Saha}. Inset shows dependence of ratio $n_T/n_e$ on the temperature of the electron gas. The parameters of calculation are $G_X = 4.5\cdot 10^{10}$~cm$^{-2}$ps$^{-1}$, $\gamma = 0.1$~cm$^{2}$s$^{-1}$, $n_c = 10^{12}$~cm$^{-2}$, $\tau_X = \tau_T = 50$~ps, $\beta = w_0\frac{k_BT}{E_T}e^{-\frac{E_T}{k_BT}}$, with $w_0 = 1.16\cdot10^3$~cm$^{2}$s$^{-1}$, $E_T = 25~\text{meV}$~\cite{venanzi2024ultrafast, shentsev2025terahertz,ayari2020phonon}.}
    \label{fig1}
\end{figure*}

For efficient heating of the electron gas, we propose using a low-frequency or static electromagnetic field satisfying $\omega\tau_{tr} \ll 1$, 
%($\hbar\omega \ll E_T$  and the direct exciton-trion transition with photon absorption is not realized)
where $\tau_{tr}$ is the momentum relaxation time, of resident electrons, see below. Hereafter, it is assumed that $\hbar\omega \ll E_T$ and that the direct exciton–trion transition via photon absorption does not occur. In addition, the temperature is assumed to be much lower than the exciton binding energy, so thermal dissociation of excitons can be neglected. Thus, the dynamic equations for the components of the system are as follows:
\begin{subequations}
    \label{syst}
    \begin{align}
    &\frac{d n_e}{d t} = -\gamma n_en_X+\frac{n_T}{\tau_{T}} + \beta n_en_T, \\
    &\frac{d n_X}{d t} =G_X  - \gamma n_en_X-\frac{n_X}{\tau_{X}} + \beta n_en_T,\\
    &\frac{d n_T}{d t} = \gamma n_en_X-\frac{n_T}{\tau_{T}} - \beta n_en_T,\\
    &\label{systd} C\frac{d T}{d t} = Q - Q_l.
    \end{align}
\end{subequations}
Here $n_e$ denotes the density of electrons not bound into trions, $G_X$, see Fig.~\ref{fig1}(d), is the rate of photogeneration of the exciton density, $T$ is the temperature of electron gas—assumed, for simplicity, to be common to all components—and $C$ is the heat capacity of the electron gas per unit area, $\tau_X, \tau_T$, see Fig.~\ref{fig1}(b, d), are lifetimes of exciton and trion, respectively. $Q$ is the Drude absorption power density, $Q_l$ is the energy loss rate density by electron gas. 
%The process with the rate $\gamma n_e n_X$, depicted in Fig.~\ref{fig1}(c), represents the pairing of excitons and electrons into trions—a mechanism most likely mediated by optical phonon emission~\cite{ayari2020phonon}.
The process with rate $\gamma n_e n_X$, depicted in Fig.~\ref{fig1}(c), represents the pairing of excitons and electrons into trions, most likely due to the emission of optical phonons~\cite{ayari2020phonon}. The term $\beta n_e n_T$, shown in Fig.~\ref{fig1}(a), describes the impact dissociation of trions~\cite{shentsev2025terahertz}, where $\beta(T) \sim \exp{(-E_T/k_BT)}$ is exponentially activated with increasing temperature, as dissociation occurs via collisions with particles whose energy is on the order of the trion binding energy.
%The process $\gamma n_e n_X$, see Fig.~\ref{fig1}(c) is the pairing of excitons and electrons into trions, a mechanism most likely mediated by optical phonon emission~\cite{ayari2020phonon}. The term $\beta n_e n_T$, see Fig.~\ref{fig1}(a), describes the impact dissociation of trions~\cite{shentsev2025terahertz}, where $\beta(T) \sim \exp{(-E_T/k_BT)}$ is exponentially activated with increasing temperature, as dissociation occurs via collisions with particles whose energy is on the order of the trion binding energy $E_T$. 
%In system~\eqref{syst} we assume that the temperature is too low for the ionization of excitons.
%\addSasha{In the system ~\eqref{syst} we assume that the temperature is insufficient for the ionization of excitons  and that the frequency of the absorbed electromagnetic field is too low for a direct trion-exciton transition.}

The densities of all components of the system~\eqref{syst} are related by simple linear expressions:
\begin{subequations}
    \label{eq:lin}
    \begin{align}
    &n_e+n_T = n_c, \\
    &\frac{n_X}{\tau_X} + \frac{n_T}{\tau_T} = G_X,
    \end{align}
\end{subequations}
where the first equation corresponds to the conservation of resident charge carriers in the conduction band, and the second connects the equilibrium densities of excitons $n_X$ and trions $n_T$ to the rate of exciton photogeneration $G_X$. In the steady state, the rate of electron binding into trions is compensated by impact dissociation and radiative decay of trions
\begin{equation}
    \label{eq:st:st}
    \gamma n_X n_e = \beta n_T n_e + \frac{n_T}{\tau_T}.
\end{equation}
Thus, a stationary point, see Fig.\ref{fig1}, for Eqs.~\eqref{syst} at a given temperature is found by solving a quadratic equation obtained from the Eqs.~\eqref{eq:lin} and \eqref{eq:st:st}
\begin{multline}
    \label{qudeq}
\left(\beta+\gamma\frac{\tau_X}{\tau_T}\right)n_e^2 + \left(\gamma\tau_X\left(G_X-\frac{n_c}{\tau_T}\right) +\frac{1}{\tau_T}-\beta n_c\right)n_e\\-\frac{n_c}{\tau_T} = 0,
\end{multline}
that has only one root at $n_e,n_T,n_X > 0$. As can be seen in Fig.\ref{fig1}(f), the temperature dependence of the densities of free electrons $n_e$ (blue solid line) and trions $n_T$ (orange solid line) exhibits a stepwise shape. At temperatures below approximately 40 K, almost all charge carriers bind with excitons to form trions. With a further increase in temperature, a sharp rise in the impact dissociation of trions occurs, and the equilibrium concentration of trions rapidly decreases to zero.

%Although our system is not in thermodynamic equilibrium with the lattice, 
It is interesting to compare calculated from Eqs.~\eqref{syst} steady-state concentrations with those predicted by the Saha equation:
\begin{equation}
    \label{eq:Saha}
    \frac{n_en_X}{n_T} = \frac{M_eM_X}{M_T}\times\frac{k_BT}{\pi \hbar^2}e^{-\frac{E_T}{k_BT}},
\end{equation}
where $M_e$, $M_X$, and $M_T$ are the masses of the electron, exciton, and trion, respectively. In addition to the fact that our system is not in thermodynamic equilibrium, the Saha equation Eq.~\eqref{eq:Saha}, the solution of which is represented by the dashed lines in Fig.~\ref{fig1}(f), does not take into account the photogeneration and radiative recombination of excitons and trions. While the solutions of Eq.~\eqref{qudeq} and Eq.~\eqref{eq:Saha} are qualitatively similar, their differences related to the, strictly speaking, non-equilibrium nature of Eqs.~\eqref{syst} vs. equilibrium Eq.~\eqref{eq:Saha} are clearly visible in Fig.~\ref{fig1}(f).
%Together, these factors make it not well suited to our case.

The approach used to solve Eq.~\eqref{qudeq} allows the coefficients of system~\eqref{syst} to be temperature-dependent~\cite{ayari2020phonon, palummo2015exciton} but does not account for Fermi or Bose correlations between particles. This is insignificant at the low concentrations ($\lesssim 10^{12}$ cm$^{-2}$) considered here. At higher particle densities, such correlations can introduce a dependence of the coefficients on the concentration of the system components, meaning that Eq.~\eqref{qudeq} would no longer have such a simple solution. On the other hand, one can expect that such a sharp change in equilibrium concentrations, shown in Fig.~\ref{fig1}(f), will also occur in this case.

\section{\label{sec:HC}Heating and cooling of electrons and trions}

%Within the quasi-particle approach, the Drude absorption power density $Q$ per unit area is given by
Within the quasiparticle approach, the low-frequency electric current density ($\omega\tau_{tr} \ll 1$) consists of contributions from free electrons and trions:
\begin{equation}
    \label{eq:current}
    j = E\times \left(\frac{e^2n_e\tau_{e,tr}}{M_e}+\frac{e^2n_T\tau_{T,tr}}{M_T}\right),
\end{equation}
where $E^2 = \frac{4\pi}{c}I$ is the root mean square field strength in the sample when irradiated with an electromagnetic field of intensity $I$, and $e$ is the charge of an electron. Consequently, the Drude absorption power density $Q$ is therefore likewise composed of two contributions
\begin{subequations}
    \label{eq:heat}
    \begin{align}
    &Q = E\times j = Q_e + Q_T, \\
    &Q_e = E^2\times\frac{e^2n_e\tau_{e,tr}}{M_e},\\
    &Q_T = E^2\times\frac{e^2n_T\tau_{T,tr}}{M_T}.
    \end{align}
\end{subequations}
In the case of scattering by point defects, the momentum relaxation time is $\tau^{-1}_{i,tr} = nu^2_iM_i/\hbar^3$, where $n$ is the defects density and $u_i$ is the effective strength of the defects’ potential. Here and below, subscript  $i = e, T$ denotes electrons and trions, respectively. Taking this into account, the ratio of the absorption power of electrons and trions, $r_D$, has the form
\begin{equation}
    \label{eq:rD}
    r_{D}^{-1} = \frac{Q_e/n_e}{Q_T/n_T} = \left(\frac{M_Tu_T}{M_eu_e}\right)^2   \sim 10,
\end{equation}
where we consider $M_T \approx 3M_e$ and $u_T \approx u_e$. In the case the system is heated by a low-frequency field ($\omega\tau_{tr} \ll 1$), the role of longitudinal acoustic and optical phonons in Drude absorption is reduced to a renormalization of the momentum relaxation time $\tau_{\text{tr}}$.
%and the effective masses $M_e$ and $M_T$. 
For simplicity, we neglect these effects; the role of phonons in energy relaxation will be discussed below.
 
The kinetic energy loss rate $Q_l$ of the electron gas consists of the following components:
\begin{equation}
\label{eq:cool}
    Q_l = Q_{LA} + Q_O + Q_I.
\end{equation}
Here $Q_{LA}$ is relaxation via longitudinal long-wave acoustic phonons, $Q_{O}$ is the relaxation via optical phonons, and $Q_I$ denotes the energy loss due to trion impact dissociation. Given that the optical phonon energy in TMDC MLs is approximately equal to the trion binding energy $E_T$~\cite{ayari2020phonon}, we neglect any temperature change due to the process of trion formation (characterized by the rate $\gamma n_e n_X$).
%We neglect the heating or cooling associated with the formation of trions from excitons and electrons  (characterized by the rate $\gamma n_e n_X$), as this process is likely accompanied by the emission of an optical phonon with an energy approximately equal to the trion binding energy $E_T$~\cite{ayari2020phonon}. 

The energy relaxation rate via longitudinal acoustic phonons, $Q_{LA}$, is given by
\begin{equation}
    Q_{LA} = \left(\frac{n_e}{\tau_{e,LA}}+\frac{n_X}{\tau_{X,LA}}+\frac{n_T}{\tau_{T,LA}}\right)k_B(T-T_l). 
\end{equation}
Here $T_l$ is the lattice temperature and 
$
    \tau_{i, LA}^{-1} = 2M_i^2|\Xi_i|^2/\rho\hbar^3
$
is a time of energetic relaxation, where $\Xi_i$ is the deformation potential and $\rho$ is the two-dimensional mass density of the crystal. Thus, the energy relaxation rate due to acoustic phonons for trions and electrons is related as follows:
\begin{equation}
    \label{eq:rLA}
    r_{LA} = \frac{Q_{T,LA}/n_T}{Q_{e, LA}/n_e} = \left(\frac{M_T\Xi_T}{M_e\Xi_e}\right)^2\sim 10,
\end{equation}
where we also assume that $\Xi_T \approx \Xi_e$. In general, the ratio of the corresponding constants for electrons and trions—such as $u$, $\Xi$, and so on—depends on the material~\cite{ayari2020phonon, phuc2018tuning, zollner2019strain, shree2018observation}. Therefore, the difference in absorption $r_D$ and cooling $r_{LA}$ between electrons and trions can be either greater or smaller.

At electron gas temperatures of about 50 K and above, cooling via optical phonons $Q_O$ becomes significant~\cite{venanzi2021terahertz,PhysRevB.90.165436}. The corresponding energy relaxation rate is given by
\begin{multline}
   \label{eq:QO}
   Q_O = \left(n_e|D_e^{(0)}|^2M_e + n_X|D_X^{(0)}|^2M_X+n_T|D_T^{(0)}|^2M_T\right)\\
   \times \frac{e^{-\frac{\hbar\omega_{O}}{k_BT}}}{2\rho\hbar}, 
\end{multline}
where $D^{(0)}_i$ is the deformation potential responsible for the interaction of the corresponding particle with optical phonons. This expression is valid specifically for interactions with a homopolar (HP) optical phonon; however, the essential exponential dependence also holds in the general case.

The energy loss due to trion impact dissociation is described by
\begin{equation}
    \label{eq:impact}
    Q_I =  E_T\times\beta(T) n_e n_T.
\end{equation}
Although the total energy of the electron gas is conserved in this process, the trion binding energy $E_T$ is subsequently transferred to the lattice via optical phonon emission $\hbar\omega_O$.
Furthermore, since in the Boltzmann limit the system energy is $CT$, where $C = k_B(n_e+n_X+n_T)$, the left-hand side of Eq.~\eqref{systd} must contain a term of the form $\frac{dC}{dT}T$, related to the temperature change due to energy sharing between particles. We neglect this term since it scales as $\sim k_BT\times\beta(T)n_en_T$ and provides only a correction to Eq.~\eqref{eq:impact} of order no larger than $k_BT/E_T$.

Thus, from Eqs.~\eqref{eq:rD} and~\eqref{eq:rLA} it is evident that the rates of absorption and cooling due to electrons and trions can differ by an order of magnitude, as a consequence of the large trion effective mass $M_T \approx 3M_e$. Thus, the conductivity of the system, as well as its absorption and cooling characteristics, strongly depend on the ratio between free electrons and trions and can therefore be used to characterize the state of the system. As will be shown below, this can lead to significant instability of the system upon variation of the heating intensity and to the coexistence of two steady states.

\section{\label{sec:Bif}Bistability of the system under heating}

The relation between the steady-state intensity intensity $I = \frac{c}{4\pi} E^2$ as a function of electron gas temperature $T$ and the exciton photogeneration rate $G_X$ is obtained by equating absorption power~\eqref{eq:heat} to the energy loss rate~\eqref{eq:cool}
\begin{equation}
\label{Inten}
    I(T, G_X) = \frac{Q_l(n_e,n_X,n_T,T)}{\frac{4\pi}{c}\frac{e^2\tau_{e,tr}}{M_e}(n_e+r_Dn_T)},
\end{equation}
\begin{figure}[t]
    \centering
    \includegraphics[width=\linewidth]{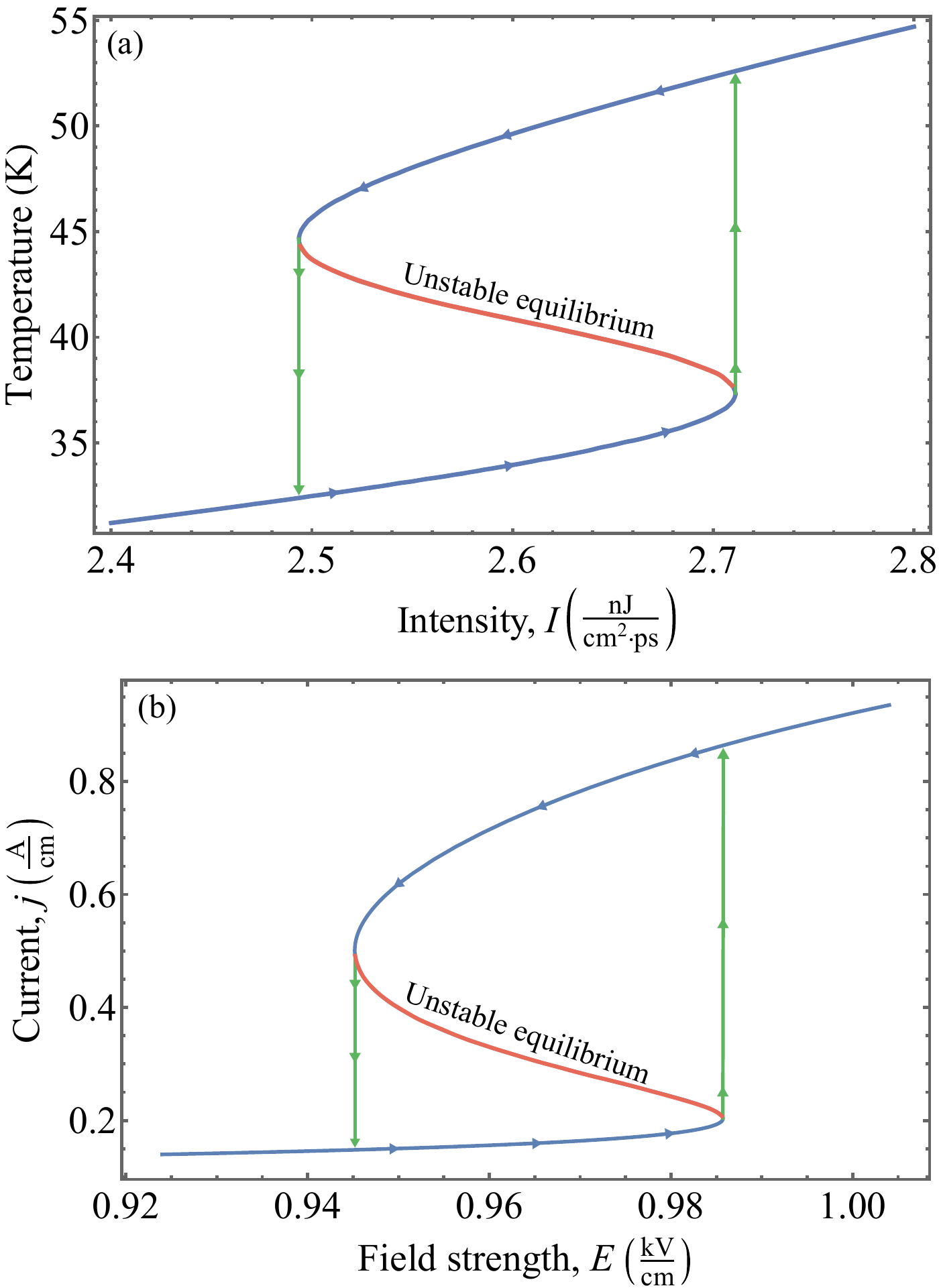}
    \caption{(a) Dependence of electron gas temperature on the heating intensity $I$. (b) Dependence of electric current density $j$ on the applied electromagnetic field strength $E$ in the sample. Arrows indicate the direction of traversal on the hysteresis loop. The parameters of calculation are $\tau_{e,tr} = 2~$ps, $\tau_{e,LA} = 30~$ps, $r_D = 0$, $r_{LA} = 10$, $D_e^{(0)} = D_h^{(0)} = D_T^{(0)} = 5.2\cdot 10^8~$eV$\cdot$cm$^{-1}$, $D_X^{(0)} = 0$, $\rho = 4.46\cdot10^{-7}~$g$\cdot$cm$^{-2}$, $E_T = 25~$meV, $\hbar\omega_O = 30.3~$ meV~\cite{ayari2020phonon}, other parameters are the same as ones used for the calculations for Fig.~\ref{fig1}.}
    \label{fig2}
\end{figure}
\begin{figure*}[t]
    \centering    \includegraphics[width=\textwidth]{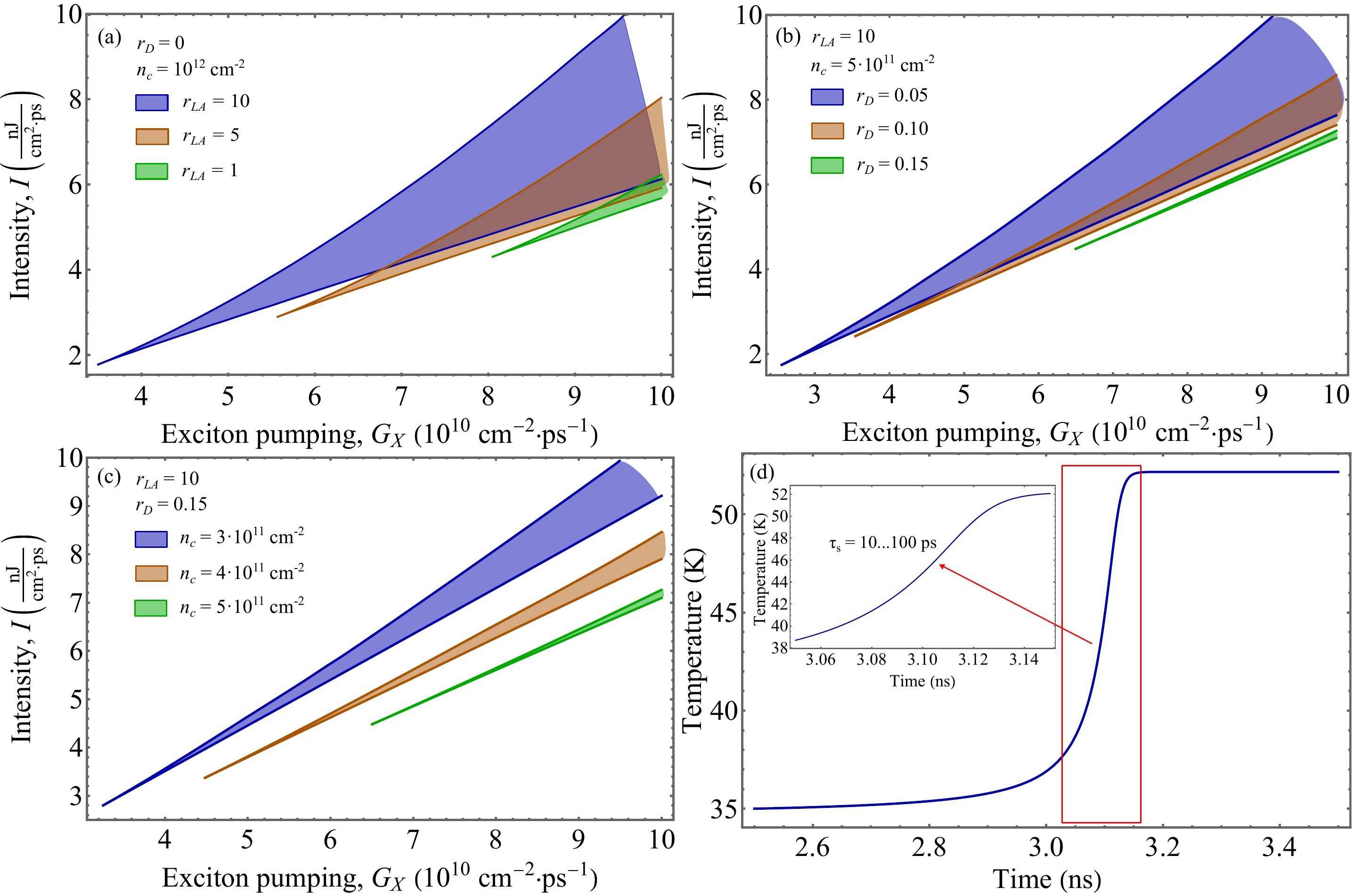}
    \caption{(a) Bifurcation region in the ($G_X$,$I$) diagram for different values of $r_{LA}$, with $r_D = 0$ and $n_c = 10^{12}$ cm$^{-2}$. (b) Same for different values $r_D$, with $r_{LA} = 10$ and $n_c = 5\cdot 10^{11}$ cm$^{-2}$, (c) Same for different values $n_c$, with $r_D = 0.15$ and $r_{LA} = 10$. (d) Temporal evolution of temperature, corresponding to the right vertical line in Fig.~\ref{fig2}, with an initial deviation from equilibrium of 0.1 K. The characteristic switching time is $\tau_s \sim 10...100$ ps.}
    \label{fig3}
\end{figure*} 
where $n_e,n_X,n_T$ are determined from Eqs.~\eqref{qudeq} and \eqref{eq:lin}.  As shown in the inset of Fig.~\ref{fig1}, the density ratio $n_T/n_e$ exhibits a step-like dependence. Given the difference in absorption and cooling capacity between trions and electrons, this can result in a non-monotonic dependence of the intensity $I(T,G_X)$ on the electron gas temperature $T$. Thus, for a certain range of intensities $I$ or field strengths $E$, the stationary solution bifurcates into two stable branches and one unstable branch, see Fig.~\ref{fig2}. The lower and upper branches correspond to the low-temperature and high-temperature plateaus in Fig.~\ref{fig1}, respectively. States below the unstable curve—the red line in Fig.~\ref{fig2}—relax to the low-temperature branch, while those above it relax to the high-temperature branch. The green vertical lines demarcate the boundaries of the hysteresis loop for various system characteristics, including temperature, electric current, the absorbed and transmitted fractions of the external electromagnetic field, and the densities of excitons and trions.

The bifurcation region in the $(G_X, I)$ diagram is highly sensitive to system parameters, as seen in Fig.~\ref{fig3}(a--c). It narrows and shifts toward higher values of both the exciton photogeneration rate $G_X$ and the heating intensity $I$, as shown in Fig.~\ref{fig3}(a,b), with decreasing difference in absorption and cooling between trions and electrons described by the parameters $r_D$ and $r_{LA}$ introduced in Eqs.~\eqref{eq:rD} and \eqref{eq:rLA}
This is a result of the fact that the loss of kinetic energy during trion impact dissociation $\beta(T) n_e n_T$, sharply increases with temperature, making the $I(T)$ dependence more monotonic. Consequently, as the differences in thermal behavior between resident charge carriers and trions diminish, the bifurcation region shifts toward a larger $n_X/n_c$ ratio, where the contribution of trion impact dissociation to $Q_l$ is less significant, see Fig.~\ref{fig3}(c). In order to get more detailed insight into the physics of bistability we consider below a simplified limiting case where $n_c \ll n_X$.
\subsection{\label{sec:Lowdenc} Low electron density regime}
As can be seen in Fig.~\ref{fig3}(c), the bifurcation effect is more pronounced at low densities of resident charge carriers $n_c\ll n_X$. In this regime, the equilibrium electron density can be written as
\begin{equation}
\label{nelow}
    n_e = 
    \begin{cases}
    \frac{n_c}{1 + \gamma\tau_T n_X}~\text{at}~\beta \ll \gamma n_X/n_c,\\
    n_c - \frac{\gamma n_X}{\beta}~\text{at}~\beta \gg \gamma n_X/n_c,
    \end{cases}
\end{equation}
where we consider the exciton density to be constant, $n_X = G_X\tau_X - \tau_Xn_T/\tau_T \approx G_X\tau_X$. The temperature on the lower (L) branche is described by
\begin{equation}
    \label{eq:Lowtemp}
    T_L = T_l + E^2\times \theta_L \left(\frac{1-r_D}{1+\gamma\tau_Tn_X}+r_D\right),
\end{equation}
where $\theta_L$ is determined by the system parameters
\begin{multline}
    \theta_L = \left(\frac{k_Bn_X}{\tau_{X,LA}}+\frac{r_{LA}k_Bn_c}{\tau_{e,LA}}\times\frac{\gamma\tau_T n_X}{1 + \gamma\tau_T n_X}
\right)^{-1}\\\times \frac{e^2n_c\tau_{e,tr}}{M_e}.
\end{multline}
Here we neglect the contribution of impact dissociation $Q_I$ and optical phonons $Q_O$ to cooling on the lower hysteresis branch, since $n_c \ll n_X$ and the deformation potential for excitons $D_X^{(0)}\sim 0.1 D_e^{(0)}$ is small in TMDC MLs~\cite{ayari2020phonon}. Therefore, its contribution is insignificant at these temperatures. At the same time, we cannot neglect the trion contribution to $Q_{LA}$, because $r_{LA}n_c$ and $n_X$ can be of the same order for large $r_{LA}$. If we disregard the optical phonon contribution, then for the upper (U) branch we obtain the following equation
\begin{equation}
    \label{eq:Uptemp}
    T_U = T_l + E^2\times \theta_U \left(1 - (1-r_D)\frac{\gamma n_X}{\beta(T_U) n_c}\right),
\end{equation}
where $\theta_U$ has the form
\begin{equation}
    \theta_U = \frac{\tau_{X,LA}}{k_Bn_X}\times\frac{e^2n_c\tau_{e,tr}}{M_e}.
\end{equation}

This approach gives predict the following expression for the temperature difference between the branches: $T_U - T_L \sim \theta_U \times E^2$. In practice, while reducing $n_c$ or increasing $n_X$ diminishes the significance of exponentially activated processes on the lower branch, it thereby shifts $T_U$ to higher temperatures, where these processes remain relevant, see Fig.~\ref{fig4}(a). As evident from the Fig.~\ref{fig4}, even at $n_X/n_c = 12.5$, cooling via optical phonon $Q_O$ is the dominant process on the upper branch. Note that cooling is primarily via electrons and trions, with the exciton contribution being insignificant at $D_X = 0.1 D_e$. Consequently, the optical phonon contribution~\eqref{eq:QO} must be included in the upper branch analysis, as it largely governs the branch's location. The slowdown in the growth of cooling via acoustic phonons $Q_{LA}$ with increasing temperature is due to a decrease in the proportion of trions, whose energy relaxation occurs faster than that of free electrons. Meanwhile, the impact dissociation term $Q_I$, while still appreciable at this $n_c$, diminishes as the resident charge carrier density $n_c$ further decreases, since this process scales as $Q_I \sim n_c^2$.

\subsection{\label{sec:Switch} Switching time}
\begin{figure}[t]
    \centering
    \includegraphics[width=\linewidth]{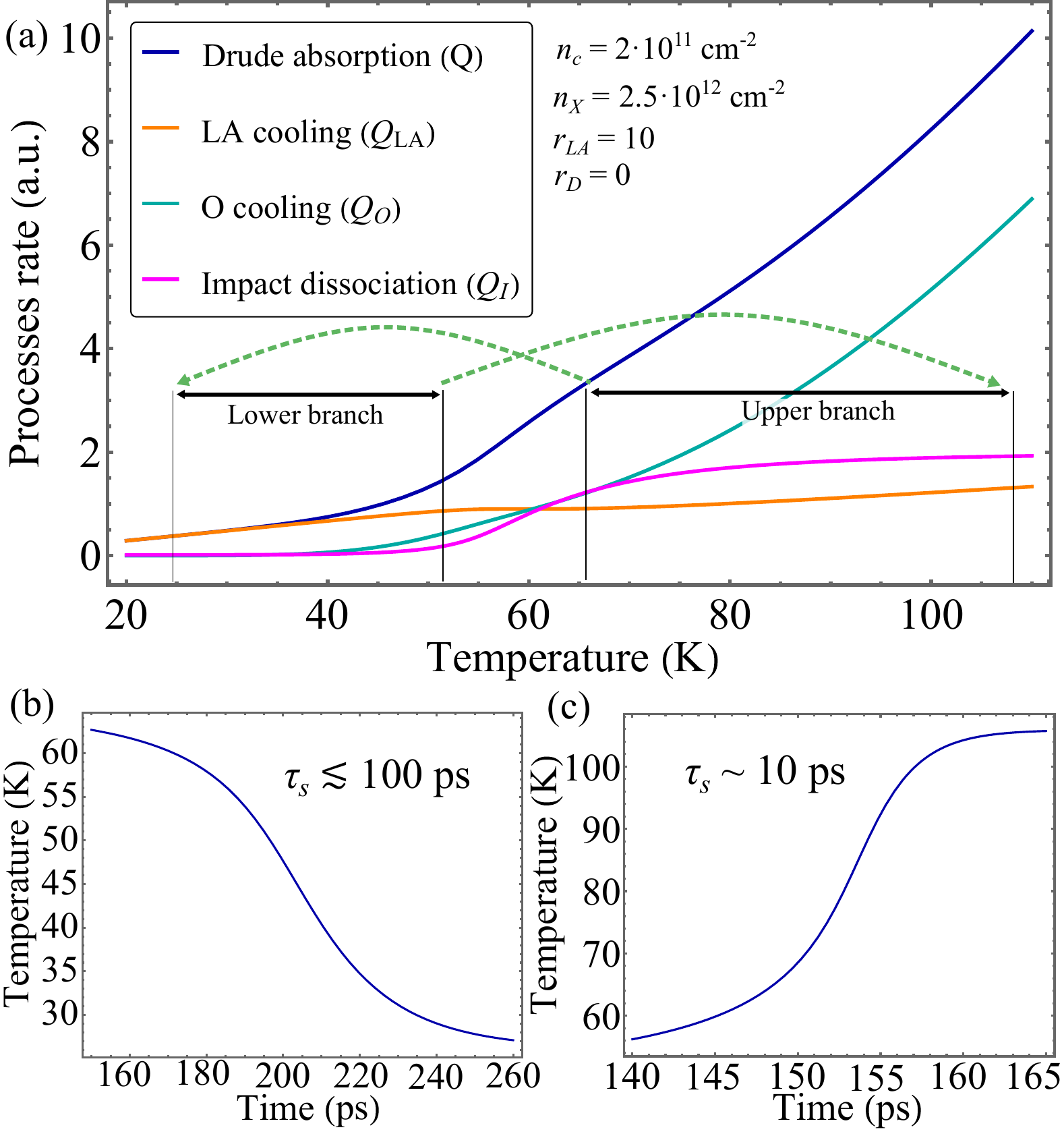}
    \caption{(a) Dependence of the thermal process power (a.u.) on electron gas temperature (K). Black arrows illustrate the temperature ranges where bistability exists. Green dashed lines correspond to switching between branches at the boundaries of the hysteresis loop, as indicated by the green vertical lines in Fig.~\ref{fig2}. (b) Temporal evolution of temperature corresponding to the left green dotted arrow from (a), namely, to the switching process with decreasing temperature. The characteristic switching time is $\tau_s \lesssim 100$ ps. (c) Temporal evolution of temperature corresponding to the right green dotted arrow from (a), i.e., to the switching process with increasing temperature. The characteristic switching time is $\tau_s \sim 10$ ps, which is considerably shorter than that in (b). The parameters of calculation are $n_c = 2\cdot 10^{11}$~cm$^{-2}$, $G_X = 5\cdot 10^{10}$~cm$^{-2}$ps$^{-1}$~cm$^{-2}$, $r_{LA} = 10$, $r_D = 0$, the initial deviation of temperature from equilibrium for (b,c) is $\lesssim 1$~К, $D_X^{(0)} = 5.2\cdot 10^7~$eV$\cdot$cm$^{-1}$, other parameters are the same as ones used for the calculations for Fig.~\ref{fig1} and Fig.~\ref{fig2}.}
    \label{fig4}
\end{figure}

The switching process corresponding to the green vertical lines in Fig.~\ref{fig2} (and dashed green lines in Fig.~\ref{fig4}(a)) involves changes in both temperature and the electron-to-trion density ratio. Here by switching time we refer to the characteristic time in significant non-equilibrium, rather then the characteristic time of exiting or establishing equilibrium, which determines the tails in Fig.~\ref{fig3}(d), which will be discussed later. For $n_X \sim n_c \sim 10^{12}$~cm$^{-2}$, the characteristic switching time $\tau_s$ is on the order of tens to hundreds of picoseconds, as shown in Fig.~\ref{fig3}(d).  One can naively separate the switching process into conversion and heating processes. The characteristic time of conversion processes $\tau_C$ can be expressed as the inverse of the transition rate per electron
\begin{equation}
    \tau_C \sim \frac{1}{\gamma n_X} \sim \frac{1}{\beta n_T}.
\end{equation}
Here, the rate of trion formation $\gamma n_X$ is of the same order of magnitude as the impact dissociation rate $\beta n_T$. This follows from the fact that the radiative term in Eq.~\eqref{eq:st:st}, normalized to the concentration of unpaired electrons $n_e$, has the form $\frac{1}{\tau_T}\frac{n_T}{n_e}$ and decreases rapidly over almost the entire temperature range under consideration, as shown in the inset of Fig.~\ref{fig1}(f), except near the onset of the lower branch.
%This holds over most of the temperature range under consideration, except near the onset of the low-temperature branch. 
With increasing exciton density, the characteristic conversion time decreases, which can be attributed to enhanced nonlinear terms that lead to faster dynamics in the system. The characteristic heating time $\tau_H$ is
\begin{equation}
    \label{eq:tau:H}
     \tau_H \sim \frac{ k_B(n_c+n_X)\times (T_U-T_L)}{\frac{4\pi}{c}I \times \frac{n_ce^2 \tau_{e,tr}}{M_e}},
\end{equation}
where, for this estimate, the heat capacity is calculated in the Boltzmann limit. In the limit $n_c \ll n_X$, the switching dynamics are asymmetric, as shown in Fig.~\ref{fig4}(b,c): Switching with increasing temperature, i.e., from from the lower temperature to the higher temperature state, occurs notably faster ($\tau_s \sim 10$ ps) than with decreasing temperature ($\tau_s \lesssim 100$ ps). For $T_U-T_L$ from the preceding section, Eq.~\eqref{eq:tau:H} reduces to the simple form $\tau_H \sim \tau_{X,LA}\sim 50$ ps.
%If we use the results for $T_U-T_L$ from the preceding section, Eq.~\eqref{eq:tau:H} reduces to the simple form $\tau_H \sim \tau_{X,LA}\sim 50$ ps, which does not explain this asymmetry.
%This contradiction is resolved by noting that 
The asymmetry arises because including cooling via optical phonons in Eq.~\eqref{eq:Uptemp}—as discussed in Sec.~\ref{sec:Lowdenc}— significantly limits the growth of the temperature difference $T_U-T_L$ corresponding to the switching with increasing temperature (right dashed green line in Fig.~\ref{fig4}(a)) and, to a lesser extent, affects switching with decreasing temperature (left dashed green line in Fig.~\ref{fig4}(a)). This, in turn, reduces the characteristic thermal switching time $\tau_H$: the system's response rate to temperature changes increases with intensity $I\sim E^2$, while the temperature range $T_U-T_L$ varies only slightly. This decrease occurs asymmetrically, which explains the behavior observed in Fig.~\ref{fig4}(b,c). Moreover, since an increase in the exciton density $n_X$ leads to higher intensity, it consequently reduces both $\tau_H$ and the overall switching time $\tau_s = \tau_C + \tau_H$.

As for the tails, the departure from equilibrium (left tail in Fig.~\ref{fig3}(f)) always occurs more slower than its subsequent establishment (right tail in Fig.~\ref{fig3}(f)). This can be described as follows: Let the temperature evolution over time near the equilibrium temperature be given by
\begin{equation}
    \frac{d \Delta T}{d t} = a\Delta T + b (\Delta T)^2,
\end{equation}
where $\Delta T$ denotes the deviation of the temperature from the equilibrium one. Here, a stable equilibrium corresponds to $a < 0$, while an unstable equilibrium corresponds to $a > 0$. Since at the boundaries (green vertical lines) of the hysteresis loop, see Fig.~\ref{fig2}, the switching between branches begins at the intersection of stable(blue lines) and unstable(red line) equilibrium, where $a = 0$  (with $b > 0$ for the right boundary and $b < 0$ for the left), it follows that the time to exit equilibrium scales as $\sim (\Delta T)^{-1}$, whereas the time of its subsequent establishment scales as $\sim \ln(\Delta T)$.

\section{\label{sec:Concl}Conclusion}
In this work, we have examined the conditions for bistability arising from the strong differences in the absorption and cooling capacities of electrons and trions in atomically thin TMDCs. In manifests itself as a reduction in the conductivity of the electron gas due to the formation of trions, wich are the bound states consisting of an electron and a heavier, weakly field-interacting exciton. This effect, combined with the step-like dependence of the equilibrium electron-to-trion density ratio on temperature, leads to bistability and hysteresis in various system quantities, namely temperature, electric current, fraction of the absorbed-transmitted external electromagnetic field and the intensity of exciton or trion luminescence.

We present an explicit expression for the intensity as a function of electron gas temperature, which can be non-monotonic. This non-monotonicity leads to the emergence of two steady states when the system is heated by an electromagnetic field.  The switching time between these two states is on the order of 10...100 ps and increases with increase in the ratio of the exciton density to the resident charge carrier density. Moreover, as the difference in absorption and cooling between trions and electrons decreases, the bistability region narrows and shifts toward a higher exciton photogeneration rate. 

In contrast to bulk silicon, where heating by a microwave field with presence exciton photogeneration induces self-oscillations in the density of electrons and holes not bound into excitons~\cite{ashkinadze1987self,Ashkinadze:1988aa}, such behavior does not occur in the present system.

Thus, our results demonstrate the possibility of bistability upon heating of TMDC MLs under continuous exciton photogeneration.

\acknowledgments
The author is grateful M. M. Glazov for valuable discussions and  acknowledges support of the Ministry of Science and Higher Education of the Russian Federation (project no. FFWR-2024-0017).

\newpage
%\nocite{*}
\let\itshape\upshape

\end{document}